\documentclass[prb,aps,nobalancelastpage,superscriptaddress,twocolumn,longbibliography]{revtex4-2}

\usepackage[utf8]{inputenc}
\usepackage[american,british]{babel}
\usepackage[T1]{fontenc}
\usepackage[pdftex]{graphicx}  
\usepackage{graphicx, xcolor}
\usepackage{dcolumn}
\usepackage{physics}
\usepackage{braket}
\usepackage{bm}
\usepackage{amsmath,amsthm,amssymb}
\usepackage{color}
\usepackage{verbatim}
\usepackage[normalem]{ulem}

\usepackage{hyperref}
\hypersetup{
 colorlinks=true,
 linkcolor=blue,
 anchorcolor = blue,
 citecolor = blue,
 filecolor = blue,
 urlcolor = blue 
}

\def \be {\begin{equation}} 
\def \ee {\end{equation}} 
\def \l {\left(} 
\def \r {\right)} 
\def \la {\langle} 
\def \ra {\rangle}  

\date{}

\begin{document}
 
\title{
Transport in a System with a Tower of Quantum Many-Body Scars
}
\date{\today}

\author{Gianluca Morettini}
\affiliation{Universit\'e Paris-Saclay, CNRS, LPTMS, 91405, Orsay, France.}

\author{Luca Capizzi}
\affiliation{Universit\'e Paris-Saclay, CNRS, LPTMS, 91405, Orsay, France.}

\author{Maurizio Fagotti}
\affiliation{Universit\'e Paris-Saclay, CNRS, LPTMS, 91405, Orsay, France.}

\author{Leonardo Mazza}
\affiliation{Universit\'e Paris-Saclay, CNRS, LPTMS, 91405, Orsay, France.}
\affiliation{Institut Universitaire de France, 75005, Paris, France.}

\begin{abstract}
We report the observation of unconventional transport phenomena in a spin-1 model that supports a tower of quantum many-body scars, and we discuss their properties uncovering their peculiar nature. 
In quantum many-body systems, the late-time dynamics of local observables are typically governed by conserved operators with local densities, such as energy and magnetization. In the model under investigation, however, there is an additional dynamical symmetry restricted to the subspace of the Hilbert space spanned by the quantum many-body scars. 
The latter significantly slows the decay of autocorrelation functions of certain coherent states of quantum many-body scars and is responsible for the unconventional form of transport that we detect numerically. 
We show that excited states with energy close to that of the quantum many-body scars play a crucial role in sustaining the transport. Finally, we propose a generalized eigenstate thermalization hypothesis to describe specific properties of states with energy close to the scars.
\end{abstract}

\maketitle

\section{Introduction}

Experiments on Rydberg atomic platforms~\cite{Bernier-17} have shown long-lived oscillations in the dynamics of local observables, posing a theoretical puzzle as they challenge conventional expectations of late-time thermalization.
A widely accepted mechanism underlying this phenomenon relies on the presence of rare eigenstates in the middle of the spectrum, dubbed \textit{quantum many-body scars} (QMBSs)~\cite{sm-17,Turner_2018,  Serbyn-21}, that violate the Eigenstate Thermalization Hypothesis (ETH)~\cite{berry1977,Deutsch-91,Srednicki-99, D_Alessio_2016}. Specifically, it is believed that the time evolution of a state with significant overlap with scars that are approximately equispaced in energy will exhibit oscillations. Albeit this mechanism is corroborated by numerical analyses on prototypical models, such as the PXP chain~\cite{Turner_2018, Serbyn-21}, its stability in the thermodynamic limit and the eventual fate of the oscillations is still debated~\cite{Surace-20, OmiyaMueller_2023_pra, OmiyaMueller_2023_prb, Giudici2024_prl, pal2024scarinduced, mueller2024semiclassical, kerschbaumer2024quantummany}. 

The dynamics of Rydberg-based setups continues to reveal unexpected phenomena that challenge simple theoretical explanations. In this respect, particular attention has been devoted to the transport of conserved charges in kinetically constrained models~\cite{zf-21,zbf-21, bidzhiev_2022_prl, dgvb-22,zbbf-22,Yang-22,lg-23,bls-23,bl-24, fagotti_prx_2024}. 
Ref.~\cite{ldsp-23} presents numerical evidence of superdiffusive behavior in the autocorrelation function of the energy density at infinite temperature: this is rather unexpected since superdiffusion is usually associated with non-abelian symmetries in integrable models~\cite{idgvw-21, denardis2021_prl, Bulchandani_2021, Gopalakrishnan_2024}. 
%Whether quantum many-body scars can directly influence thermodynamic transport phenomena remains an open question.

The study of the physics associated to QMBSs has benefited from the identification of simple models that are not directly related to Rydberg-atom chains where QMBSs can be found exactly at any finite size, and also be analytically described~\cite{mm-20,Moudgalya_2022,Chandran-23,mm-24}. 
Being equispaced in energy and generated by a ladder operator, these states are often called a tower of QMBSs.
In these models it is possible to propose uncorrelated product states that are coherent linear superpositions of QMBSs and which display undamped oscillatory dynamics.
Although the price to pay is that of losing a direct link with the experimental platforms where scars have been first found, the advantage is that the study of these setups is theoretically more manageable and can give an analytical intuition of the physics associated to QMBSs.

In this article, we want to study the transport properties associated to QMBSs using one of the aforementioned models supporting exactly-solvable QMBSs.
Although motivated by the study in Ref.~\cite{ldsp-23}, our work differs from it significantly in the fact that we do not consider an infinite-temperature state, but instead a coherent superposition of QMBSs.
By computing the autocorrelation function of some single-site operators that are naturally associated to QMBSs, the so-called ladder operators that generate the entire set of QMBSs, we can study its spatial and time profile, and identify whether its decay to zero is compatible with a ballistic spreading, diffusive or another type of behavior.

Specifically, we identify an unconventional mechanism of transport that is intrinsically linked to the presence of QMBSs.
% We demonstrate this in a quantum spin chain that features analytically known QMBSs. Notably, the Hilbert space spanned by these scars includes uncorrelated product states, known as coherent states, which display perfect revivals under time evolution. We focus on those states and study the autocorrelation functions of local operators as a function of space and time. 
We observe peculiar slowly decaying oscillations stemming from a dynamical symmetry within the scar subspace: this mechanism, pinpointing unconventional transport, is absent in the infinite-temperature state. 
The numerical analysis that we present shows that, in the time range accessible to our numerics, the transport is compatible with a superdiffusive scaling, and a dynamical critical exponent $z=3/2$. While a ballistic behaviour can be safely ruled out, a diffusive one is more debatable. In particular, it is not to be excluded that on time scales longer than those that we could study, the system crosses over towards a diffusive behaviour with dynamical critical exponent $z=2$.
Remarkably, the part of the many-body spectrum that is compatible with ETH mediates the transport phenomenon that we observe numerically, while the QMBSs give a vanishing contribution in the thermodynamic limit; we propose a conjecture, extending the usual ETH ansatz, to explain that.

%%%
We organize the manuscript as follows. We introduce a paradigmatic model hosting an exact tower of quantum many-body scars in Sec.~\ref{sec:model}. In Sec.~\ref{sec:autocorr} we discuss the properties of an autocorrelation function that unveil an underlying unconventional transport mechanism. A proof of the irrelevance of the scars inside the autocorrelator is provided in Sec.~\ref{sec:irrelevance}, and a conjecture to explain the unconventional transport is discussed in Sec.~\ref{sec:ETH}. We outline the conclusions in Sec.~\ref{sec:conclusion}, leaving some technical details to the Appendix.

\section{The model}\label{sec:model}

We consider a spin-1 chain described by the Hamiltonian~\cite{si-19k, Chandran-23}
\begin{multline}\label{eq:ham}
H = J \sum\limits_{j}\left( S^x_{j} S^x_{j+1} + S^y_{j} S^y_{j+1} \right) + h \sum_{j} \left( S_{j}^{z}+1 \right)\\
+ D \sum_j \left(\left( S_j^{z} \right)^2-1 \right) +J_3 \sum_{j} \left(S_{ j}^x S^x_{j+3}+S_{ j}^y S^y_{j+3}\right),
\end{multline}
with $S^a_j$ ($a=x,y,z$) being spin operators at site $j$. Together with the energy, the model conserves $\sum_j S^z_j$, the magnetization along the $z$-axis; moreover, for $J_3 =0$ it possesses a non-local $SU(2)$ symmetry~\cite{khn-03}, which is broken when $J_3 \neq 0$. We set $\hbar =1$.

Ref.~\cite{si-19k} showed that the model is non-integrable and possesses an exact tower of scars
\begin{equation}\label{eq:j_dag}
\ket{N} \propto (J^\dagger)^N \ket{\Downarrow}, \qquad J^\dagger = \frac{1}{2}\sum_j (-1)^j(S^+_j)^2,
\end{equation}
with $ N=0,\dots,L$ and $\ket{\Downarrow}$ the fully polarized state along the $z$ direction. These states are perfectly equispaced in energy since 
$
H\ket{N} = \omega N \ket{N}, $ with $  \omega = 2h.
$
Specifically, for every state $\ket{\psi}$ in the \textit{scarred subspace} $ W \equiv \text{Span}\{\ket{N}\}$,
% and $\Pi_W \equiv \sum_N \ketbra{N}{N}$ its orthogonal projector, 
it holds
\begin{equation} \label{eq:dyn_sym}
([H,J^\dagger] - \omega J^\dagger) \ket{\psi} = 0, \quad \forall \ket{\psi} \in W. 
\end{equation}
Thus, $J^\dagger$ acts as a dynamical symmetry~\cite{btj-19,mbj-20,mpz-20,Buca-22} in $W$ and will be referred to as the \emph{ladder operator}. The mechanism responsible for protecting the exact tower of scars is referred to in the literature as the \textit{restricted spectrum generating algebra} (RSGA)~\cite{mlm-20,mrb-20}.
For the infinite-size limit, with $N/L \in (0,1)$ held fixed (excluding the trivial fully-polarized states at
 $N/L=0$ or $1$),
the states $\ket{N}$ lie within the bulk of the energy spectrum and exhibit long-range spatial and temporal correlations~\cite{isx-19,dppgp-22}. 
This is the regime we focus on in this letter. For completeness, additional results on the infinite-size limit of scars are provided in Appendix~\ref{app:inf_vol_scars}.

In order to have access to the transport properties associated to QMBSs, we consider a family of coherent states $\{\ket{\zeta}\}$, parametrized by the complex number $\zeta$, and defined as follows~\cite{si-19k}
\begin{equation}
\ket{\zeta} \propto \exp\l \zeta J^\dagger\r\ket{\Downarrow}.
\end{equation}
One can readily show that, (i), the coherent states are linear superpositions of scars, namely $\ket{\zeta}\in W$; (ii), they are uncorrelated product states, since $\exp \left( \zeta J^\dagger \right) = \prod_j e^{\frac{\zeta}{2} (-1)^j \left( S^+_j\right)^2}$; and, (iii), they display perfect time revivals, since Eq.~\eqref{eq:dyn_sym} implies
$
e^{-iHt}\ket{\zeta} = \ket{e^{-i\omega t} \zeta}
$. 
Their explicit expression is $\ket{\zeta} \propto \prod_j \left( (-1)^j \zeta \ket{\uparrow}_j + \ket{\downarrow}_j \right)$, where $\ket{\uparrow}_j$ and $\ket{\downarrow}_j$ are the eigenvectors of $S_j^z$ with eigenvalues $\pm 1$.
For completeness, we mention that similar coherent states have been found in other models~\cite{is-20, Gotta_2022_Exact}, represented by matrix-product states (MPS) of finite bond dimension, and $J^\dagger$ is not necessarily a sum of single-site operators. 
%CONCLUSIONS Since most of our results rely on the SGA but not on the details of the interactions, we expect them to hold true also in these more generic situations.

\section{Autocorrelation functions and transport phenomenology}\label{sec:autocorr}

We study here the properties of the following connected correlation function
\be\label{eq:autocorr}
\begin{split}
\la \mathcal{O}^\dagger(x,t)\mathcal{O}(0,0)\ra_c \equiv 
&\bra{\zeta} \mathcal{O}^\dagger(x,t)\mathcal{O}(0,0)\ket{\zeta} \\
&-\bra{\zeta} \mathcal{O}^\dagger(x,t)\ket{\zeta}\bra{\zeta}\mathcal{O}(0,0)\ket{\zeta}, 
\end{split}
\ee
which will be simply referred to as  \textit{autocorrelator} in the following.
We have denoted by $\mathcal{O}$ a generic local operator, and by $\mathcal{O}(x,t)$ its translation by $x$ sites (to the right) in space and  by  $t$ in time in the Heisenberg picture. We remind the reader that nonequal-time correlators play a central role in linear response theory, particularly through the Kubo formula~\cite{Kubo-57}, which relates them to the effects of small perturbations.

To characterize~\eqref{eq:autocorr} we first observe that the Lieb-Robinson bound for the autocorrelator~\cite{lr-72} holds because the state $\ket{\zeta}$ is short-range correlated: hence, the autocorrelator is exponentially small in $t$ for $|x|\geq v_{\text{LR}} |t|$, with $v_{\text{LR}}$ the so-called \textit{Lieb-Robinson velocity}, and its support is concentrated inside the light-cone $|x|\leq v_{\text{LR}} |t|$.

Second, whenever $\mathcal{O}$ is the density of a conserved charge, say $[H,\sum_x \mathcal{O}(x)] =0$, the spatial integral of the autocorrelator is constant in time, meaning that
% \be\label{eq:int_autocorr}
$\sum_x \la \mathcal{O}^\dagger(x,t)\mathcal{O}\ra_c$
% \ee
does not depend on $t$ explicitly: for the model in Eq.~\eqref{eq:ham}, this is the case of the magnetization, $\mathcal{O} = S^z_j$, and of the energy, $\mathcal{O} = h_j$ with $H = \sum_j h_j $. In such a scenario, the autocorrelator spreads across the system and, at fixed $x$, it generically decays algebraically in time. In particular, at late times, one expects the following scaling behavior~\cite{Spohn-12}
\begin{equation}\label{eq:scaling}
\la \mathcal{O}^\dagger(x,t)\mathcal{O}(0,0)\ra_c \simeq \frac{1}{t^{1/z}}f\l\frac{x}{t^{1/z}}\r,
\end{equation}
with $z$ the dynamical critical exponent and $f$ a universal function associated with the universality class of the underlying transport.

In our model, a similar condition is satisfied by the ladder operator, as we explain below. Specifically, we say that $\mathcal{O}$ is conserved at wavelength $k$ and frequency $\Omega$ in the subspace $W$ whenever
\begin{equation}\label{eq:cons_freq_mom}
\left([H,\tilde{\mathcal O}(k)]-\Omega \tilde{\mathcal O}(k)
\right)\ket{\psi}=0, \quad \forall \ket{\psi} \in W
\end{equation}
holds, with $\tilde{\mathcal{O}}(k) = \sum_x e^{ikx} \mathcal{O}(x)$. In the presence of this unconventional form of transport, ansatz~\eqref{eq:scaling} for the autocorrelator then reads
\begin{equation}\label{eq:autocorr_osc}
\la\mathcal{O}^\dagger(x,t)\mathcal{O}(0,0)\ra_c \simeq \frac{1}{t^{1/z}}f\l\frac{x}{t^{1/z}}\r e^{ikx-i\Omega t},
\end{equation}
valid for (short-range correlated) states of $W$. In the specific case of the ladder operator $J^\dag$ in \eqref{eq:j_dag} we have $\mathcal{O} = (S^{+}_j)^2$, $k=\pi$, and $\Omega = \omega\equiv 2h$. 
The autocorrelator should therefore get a space-time modulation $(-1)^x e^{-i\omega t}$ for any coherent state $\ket{\zeta}$ in the scarred subspace $W$, superimposed on a smooth hydrodynamic spreading. Such a phenomenology is not expected for finite-temperature functions of magnetization and energy, which are the only global conserved charges (at $k,\Omega=0$) in the entire Hilbert space; we also remark that their dynamical exponent $z$ is not necessarily related to that of $J^\dagger$ appearing in \eqref{eq:autocorr_osc}.

%%%%FIG
\begin{figure}[t]
 \includegraphics[width=\columnwidth]{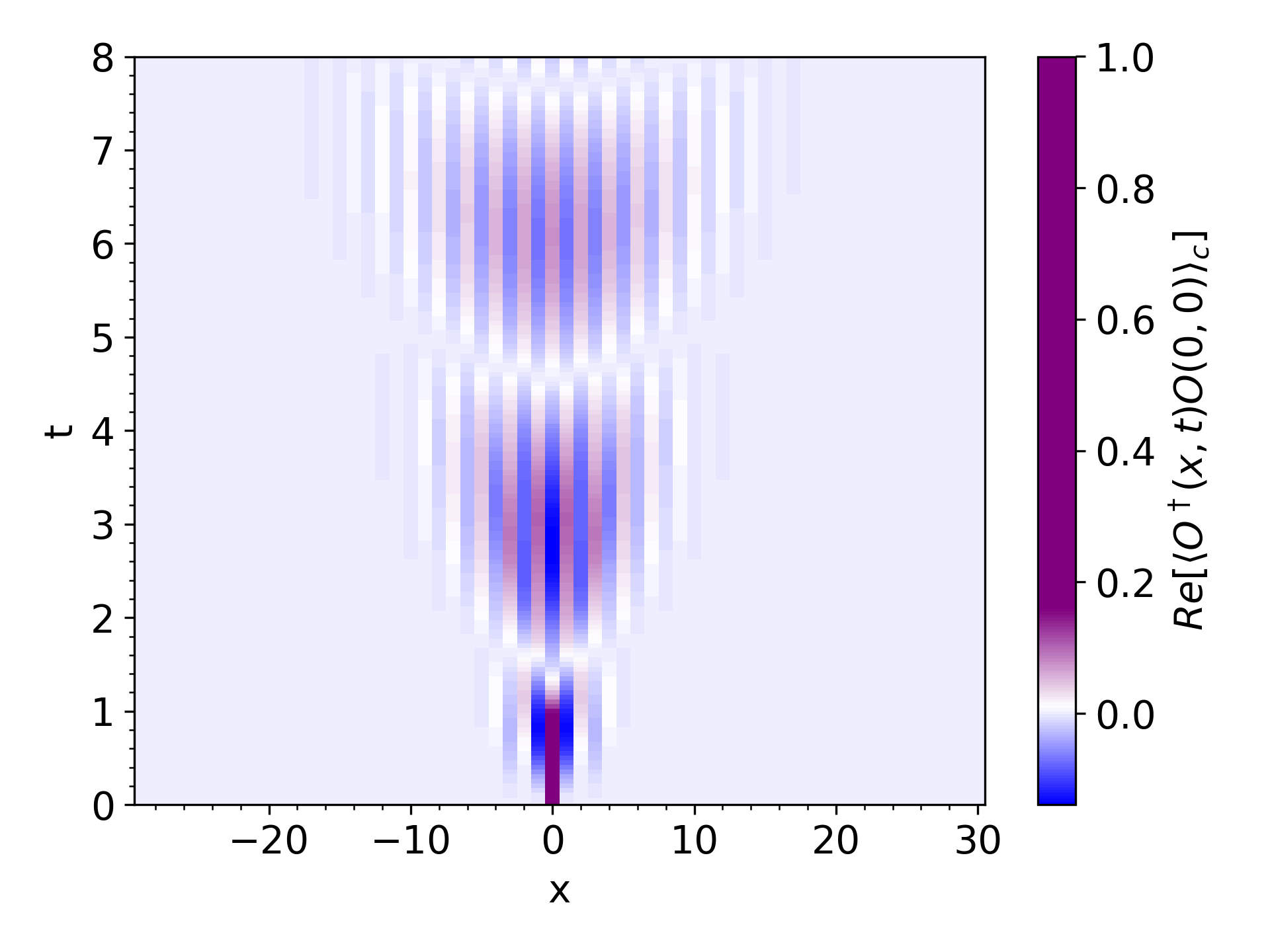}
 \caption{Real part of the autocorrelator $\langle \mathcal O^\dagger (x,t) \mathcal O(0,0) \rangle_c$ for $\mathcal{O}=(S^{+}_{L/2})^2$. 
 As time increases, a profile expands from the middle of the chain. 
 Distinct even/odd effects and oscillations with period $2\pi/\omega \simeq 6.28$ are manifest, with $\omega = 2h$. The gradient is chosen to contrast the values between $-0.15$ (blue) and $0.15$ (purple). The parameters are $J =1, D=0.1, J_3 = 0.5, h=0.5, \zeta = -i$ and the size is $L=60$.
 }
 \label{fig:xt_profile}
\end{figure}
%%%%%

We study the behavior of the autocorrelator~\eqref{eq:autocorr} using numerical tools, focusing on an open chain of even length $L=60$ and the operator $\mathcal{O} = (S^{+}_{L/2})^2$ inserted in the middle of the chain, up to times $t=8$.
The numerical simulations are performed using the ITensor library~\cite{itensor, itensor-r0.3} and we employ a maximal bond dimension $m=300$, ensuring convergence; the time-evolution scheme is the time-depending variational principle~\cite{Haegeman_2011}. 
We choose the parameters $J =1, D=0.1, J_3 = 0.5, h=0.5$, and we consider the coherent state $\ket{\zeta}$ at $\zeta = -i$; it has zero magnetization density and its energy density corresponds to infinite temperature.
The spatio-temporal behavior of the real part of $\la \mathcal{O}^\dagger(x,t)\mathcal{O}(0,0)\ra_c$ is shown in Fig.~\ref{fig:xt_profile}. 
The plot reveals a spreading profile originating from the central site that, at the considered times, is still in the bulk of the chain. %does not touch the boundary  
The pattern imprinted with $\Re e^{i( \pi x- \omega t)}$ is manifest and constitutes the first hallmark of the existence of a local ladder operator of the tower scar with a finite momentum and a finite frequency.
 This behavior demonstrates a form of persistent quantum coherence, which is discussed in greater detail below. Within the investigated time window, the correlator is exponentially close to that of an infinite system, as dictated by the Lieb-Robinson bound. This observation rules out the possibility that the emerging pattern in the figure arises from spurious effects due to finite size or boundaries (see  Appendix~\ref{App:Numerics} for additional numerical details).

%%%%FIG
\begin{figure}[t]
 \includegraphics[width=\columnwidth]{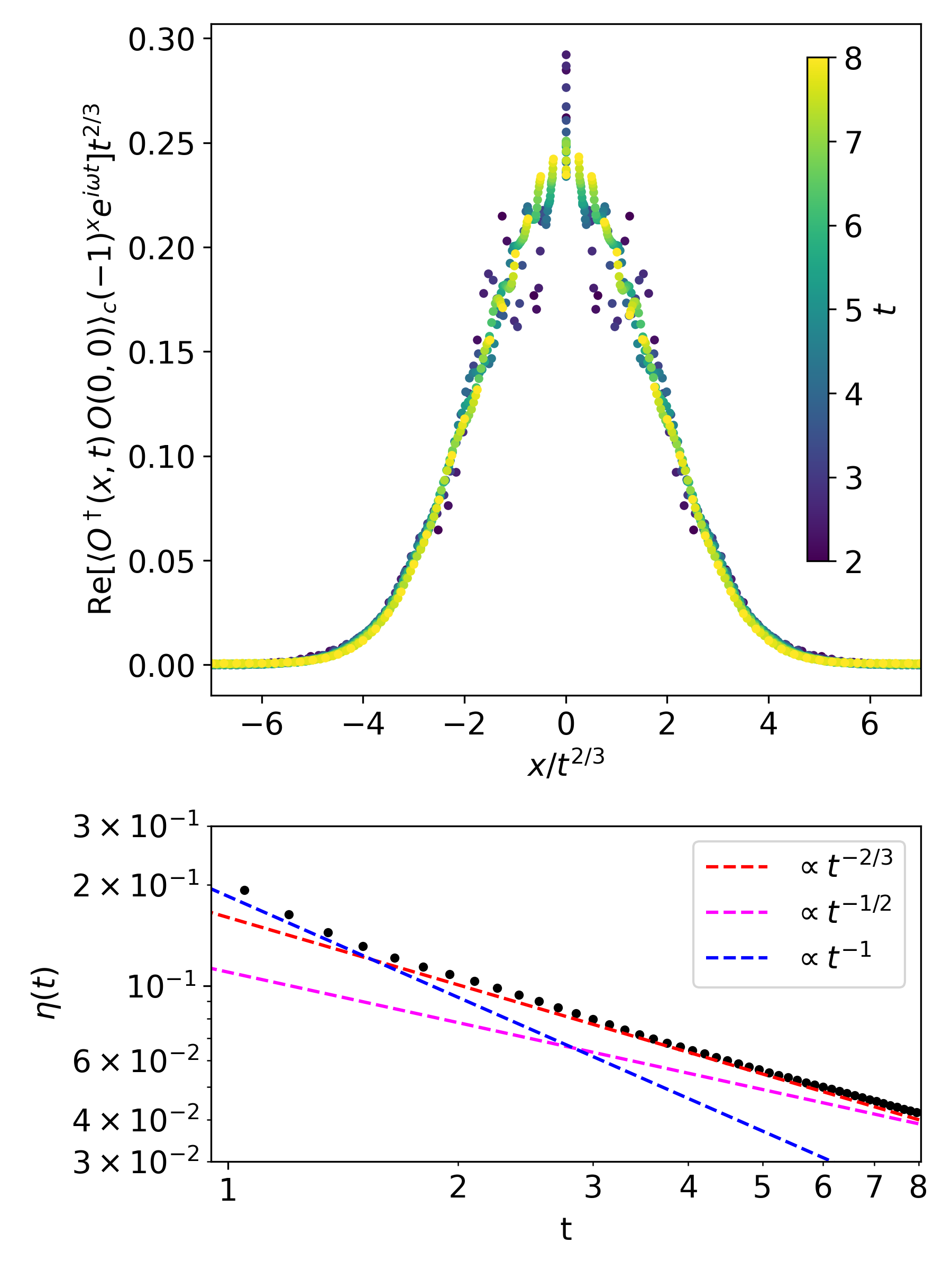}
 \caption{Top panel: Rescaled profile of $\text{Re}[\langle \mathcal{O}^\dagger(x,t)\,\mathcal{O}(0,0)\rangle_c (-1)^x e^{i\omega t}]$ using the scaling ansatz in Eq.~\eqref{eq:autocorr_osc} with $z=3/2$. We plot different values of $t$, starting from $t=2$ (purple) until $t=8$ (yellow) with time step $\Delta t = 0.25$. Bottom panel: We show the time-dependence of $\eta(t)$ defined in the text; the plotted scalings $\propto t^{-1}$, $\propto t^{-1/2}$ and $\propto t^{-2/3}$ show the compatibility of the data with the latter.}
 \label{fig:collapse}
\end{figure}
%%%%%

We now investigate the validity of Ansatz~\eqref{eq:autocorr_osc}, which is expected to hold in the thermodynamic limit for sufficiently late times. 
Our main goal is to extract the dynamical exponent $z$ from the numerical data. In Fig.~\ref{fig:collapse}~(top panel) we plot $\text{Re}
\left[ (-1)^x e^{i\omega t}\langle \mathcal O^\dagger (x,t) 
\mathcal O(0,0) \rangle_c\right] t^{1/z}$ as a function of $x/t^{1/z}$ for different values of $t$; we show the result for $z=3/2$. This choice is motivated by the behavior of the estimator $\eta(t) = \sum_x |\text{Re}
\left[ e^{i\omega t}\langle \mathcal O^\dagger (x,t) 
\mathcal O(0,0) \rangle_c\right]|^2$, whose asymptotics is $\eta(t) \sim t^{-1/z}$ assuming the ansatz~\eqref{eq:autocorr_osc}: this comes from the approximation $\eta(t)\simeq t^{-2/z}\int dx |f(x/t^{1/z})|^2 \propto t^{-1/z}$ valid at large times. In Fig.~\ref{fig:collapse} (bottom panel) we plot $\eta(t)$ alongside three reference curves $t^{-1/z}$ with $z=2,1,3/2$ corresponding to diffusive, ballistic and superdiffusive transport. \textit{The data are consistent with superdiffusion $z=3/2$ for the timescales under analysis}.

We point out that in the central region (see the top panel of Fig.~\ref{fig:collapse} at $x\simeq 0$) additional spurious oscillations are present, which we attribute to transient behavior. These effects make it difficult to confidently extract the dynamical exponent from the autocorrelator at $x=0$. However, the overall collapse appears convincing, which further supports the choice suggested by the estimator $\eta(t)$. We discuss below other possible collapses with $z=1,2$.

\begin{figure}[t]
 \includegraphics[width=\columnwidth]{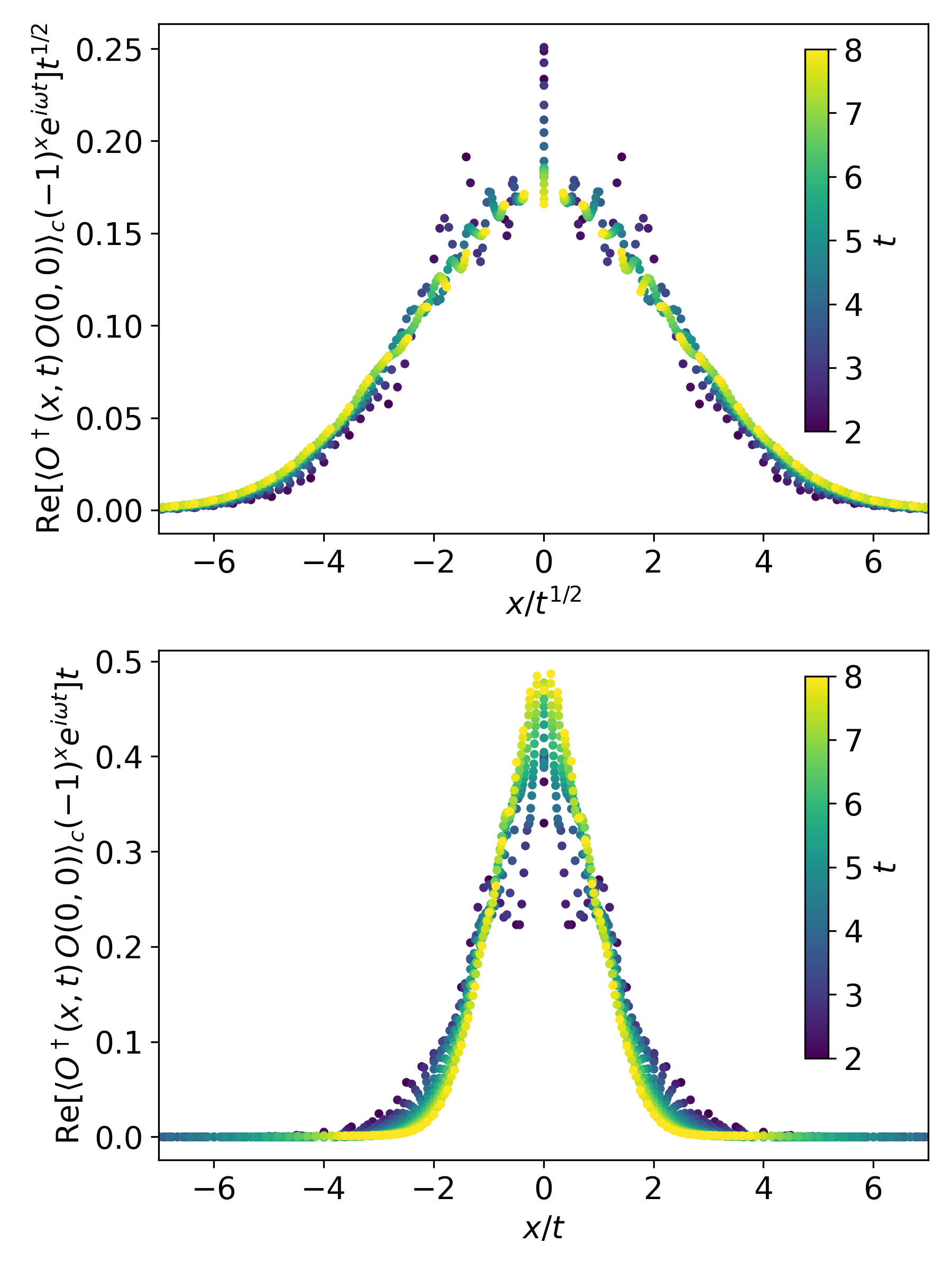}
 \caption{Rescaled profile of 
 $\text{Re}[\langle \mathcal{O}^\dagger(x,t)\,\mathcal{O}(0,0)\rangle_c (-1)^x e^{i\omega t}]$ using the scaling ansatz in Eq.~\eqref{eq:autocorr_osc} with $z=2$ (top panel) and with $z=1$ (bottom panel), by considering the (rescaled) spatial profiles at different values of $t$, starting from $t=2$ (purple) until $t=8$ (yellow) with time step $\Delta t = 0.25$. The plotted data are those presented in Fig.~\ref{fig:collapse}.
 }
 \label{fig:collapse_z1}
 \label{fig:collapse_z2}
\end{figure}

We study the profile of the autocorrelator presented in Fig.~\ref{fig:xt_profile} using the scaling ansatz in Eq.~\eqref{eq:autocorr_osc}, as a function of $x/t^{1/z}$, with $z = 2$ or $ 1$, corresponding to diffusive and ballistic transport respectively. 
In Fig.~\ref{fig:collapse_z1}(top), we show the numerical data for $\text{Re}
\left[ (-1)^xe^{i\omega t}\langle \mathcal O^\dagger (x,t) 
\mathcal O(0,0) \rangle_c\right] t^{1/z}$ with $z=2$: both the tails and the central region look incompatible with a collapse. A similar plot is presented in Fig.~\ref{fig:collapse_z2} (bottom) for $z=1$. 
Here, although the overall profile, particularly the tails, is less convincing compared to the data at $z=3/2$ in Fig.~\ref{fig:xt_profile}, the central region is not incompatible with a collapse (except at short times, where the data deviate). For this reason, we analyzed the quantity $\eta(t)$ in Fig.~\ref{fig:collapse} to discriminate between $z=2$ and $z=3/2$, tending to favor the second option.

%%%%FIG
%%%%%

We remark that the spatial integral of the autocorrelators, weighted with the oscillating factor $(-1)^xe^{i\omega t}$, is a time-independent positive constant as a consequence of the dynamical symmetry. That is why the real part of the autocorrelator, shown in Fig. \ref{fig:collapse}, gives rise to a spreading profile that is initially localized in space (at $t=0$). On the other hand, the imaginary part sums to zero and our numerical data did not follow convincing collapses compatible with the ansatz \eqref{eq:autocorr_osc}: thus, it remains an open question, beyond the aim of this work, to find an analytical theory that can explain both the real and imaginary parts from first principles.

\begin{figure}[t]
 \includegraphics[width=\columnwidth]{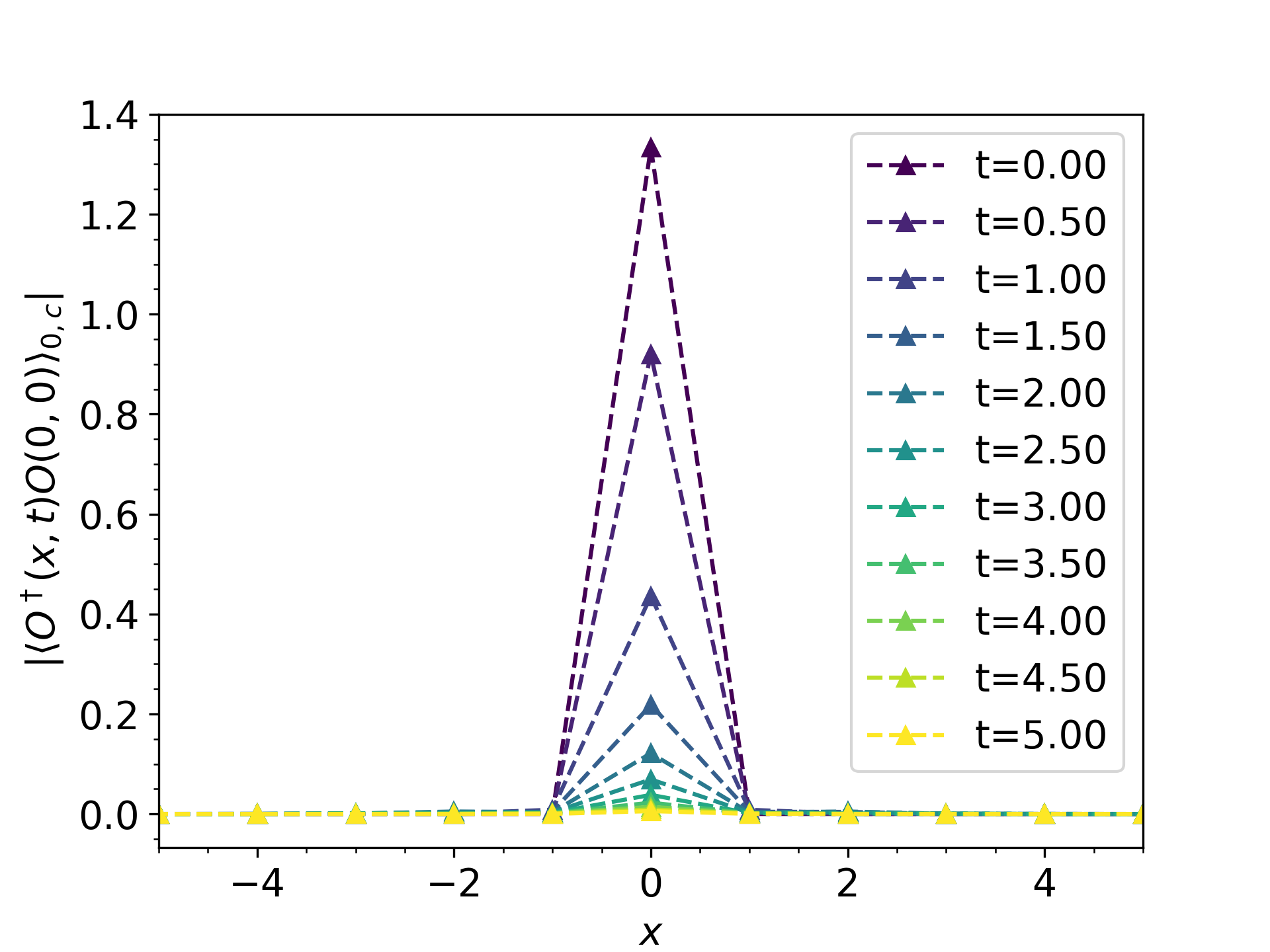}
 \caption{We show the absolute value of the autocorrelator of $\mathcal{O}=(S^{+}_{L/2})^2$ in the infinite temperature state, for $t\in[0,5]$ in a chain of length $L=60$.  A quick decay, starting from the value $4/3\simeq 1.33$ at $t=0$, is observed in time and the spatial support does not grow, differently from what we observed for the coherent state. 
 }
 \label{fig:inf_temp}
\end{figure}

Finally, the same autocorrelation function computed in the infinite temperature state has a faster decay that is not compatible with a power law. In particular, we evaluate the autocorrelator in the infinite temperature state and we denote it by $\la\mathcal{O}^\dagger(x,t)\mathcal{O}(0,0)\ra_{0,c}$. Our data in Fig.~\ref{fig:inf_temp} show a clearly different behavior compared to the coherent state reported in Fig.~\ref{fig:xt_profile}; here, a fast decay in time is observed and the support of the autocorrelator does not spread in space. This is compatible with the fact that $(S^{+}_j)^2$ does not generate any global conserved charge in the infinite temperature state, and the spatial integral of the autocorrelator is not necessarily constant in time. Specifically, the conservation of the ladder operator, expressed by~\eqref{eq:cons_freq_mom}, holds in the sector $W$ exponentially small compared to the Hilbert space. On the other hand, while we infer that scars are responsible for the results in Figs.~\ref{fig:xt_profile} and~\ref{fig:collapse}, their mere presence cannot fully explain them, as we motivate in the next section.

%%%%FIG
%%%%%

\section{Irrelevance of scar states in the autocorrelator dynamics}\label{sec:irrelevance}

In order to better understand the properties of the autocorrelator~\eqref{eq:autocorr}, we propose to express it as $ \bra{\zeta} \mathcal O^\dagger (x,t) \, \mathcal Q_\zeta \, \mathcal O(0,0) \ket{\zeta}$, where $\mathcal Q_\zeta = 1 - \ket{\zeta} \hspace{-0.1cm}\bra{\zeta}$;
this writing highlights the fact that the dynamics of the autocorrelator is determined by all states of the Hilbert space that are orthogonal to $\ket{\zeta}$.
A key result of this manuscript is that in the infinite-size limit 
\begin{equation}
\langle  \mathcal O ^\dagger (x,t)
 \mathcal O (0,0) \rangle_c 
 \simeq
 \langle \zeta | \mathcal O^\dagger (x,t) 
\, \mathcal Q_W \, \mathcal O(0,0)| \zeta \rangle 
\label{eq:conn_diff_proj},
\end{equation}
with $\mathcal Q_W = 1- \Pi_W$ and $\Pi_W = \sum_N \ket{N} \hspace{-0.1cm} \bra{N}$ the projector onto $W$: this is a stronger statement on the autocorrelator dynamics, that are thus determined only by states that are orthogonal to \textit{any} scar. The result also holds for a generic shifted and time-evolved operator $\mathcal O'(x,t)$ that is not necessarily the Hermitian conjugate of $\mathcal O$. We provide a proof of Eq.~\eqref{eq:conn_diff_proj} below, in Sec.~\ref{Subsec:Proof}.

Let us now highlight the consequences of~\eqref{eq:conn_diff_proj} in the interpretation of the data in Figs.~\ref{fig:xt_profile} and~\ref{fig:collapse}.
After introducing the energy eigenstates $\{\ket{E_i} \}$, we can write
\begin{align} \label{Eq:zeta:N:autocorr}
 \bra{\zeta} \mathcal (S^-_{ j}(t))^2 \mathcal Q_W (S^+_{ L/2})^2 \ket{\zeta} =  
 \hspace{-0.3cm}
 \sum_{\substack{N, N' \\i \text{ s.t.} \ket{E_i}\notin W} } 
 \hspace{-0.3cm}
 \bra{\zeta} N \rangle \bra{N'} \zeta \rangle \times \nonumber \\
  \times e^{- i (E_i -\omega N) t} \bra{N} \mathcal (S^-_{j})^2 \ket{E_i}
 \bra{E_i} (S^+_{L/2})^2 \ket{N'},
\end{align}
where the sum over $i$ is restricted to those eigenstates that are not in $W$.
The off-diagonal matrix elements $\bra{E_i} (S^+_j)^2 \ket{N}$ give an oscillating contribution in time at frequency $E_i - \omega N $:
the autocorrelator oscillations at frequency $\omega$ highlighted in Fig.~\ref{fig:xt_profile} must be due to one or more eigenstates whose energy is $E_i \simeq \omega (N+1)$.
It would be tempting to identify them with the QMBS $\ket{N+1}$, which has \textit{exactly} the required energy, and for which 
$ \lim_{L \to \infty}
\bra{N+1} (S^{+}_j)^2 \ket{N}  \neq 0
$
for $N/L$ fixed (the latter observation appears in Ref.~\cite{isx-19} but we derive it with different saddle-point techniques in Appendix~\ref{app:inf_vol_scars} for completeness). In summary, the scars themselves are in the kernel of $\mathcal{Q}_W$ and do not participate in the dynamics: they cannot be directly responsible for long-lived oscillations and unconventional transport.

A physical grasp on the irrelevance of the scars is obtained by considering the fact that
the scars $\ket{N}$ have momentum $k=0$ or $\pi$ for even or odd $N$.
As is intuitively clear, the spreading structure in Fig.~\ref{fig:xt_profile} must be due to states with all possible momenta $k \in[0, 2\pi)$ and in the thermodynamic limit it is not affected by the presence or absence of states with specific momenta $k=0, \pi$.

\subsection{Proof of Eq.~\eqref{eq:conn_diff_proj}}
\label{Subsec:Proof}

We start by expressing the projector $\Pi_W$ as a function of the coherent states; 
in principle, this is standard textbook material \cite{Radcliffe-71}, and it amounts to express the identity operator in terms of coherent states in a spin-$S$ representation of $SU(2)$ for large $S$ (here, $2S=L$). For completeness, we provide here a concise derivation using saddle-point techniques.

We begin by arguing that $\Pi_W$ should have the following form:
\begin{equation}\label{eq:Proj_scar}
\Pi_W \simeq L \int_{\mathbb{C}} \frac{d^2\zeta'}{\pi} F\l|\zeta'|^2\r
\ket{\zeta'} \hspace{-0.1cm} \bra{\zeta'},
\end{equation}
for some unspecified function $F$ which depends on $\zeta'$ through its modulus. 
Indeed, by taking the matrix element $\bra{N} \Pi_W \ket{N'}$ and using the expression for the scalar product $\bra{N} \zeta \rangle \propto \zeta^N$, after performing the complex-plane integral we obtain that $\bra{N} \Pi_W \ket{N'} =0$, as it should be, whenever $N \neq N'$.
The specific functional form of $F$ is identified by the requirement $\bra{\zeta}\Pi_W\ket{\zeta}=1, \;\forall \ket{\zeta}$; 
this problem is not relevant for the article, but for completeness the explicit expression $F(|\zeta|^2) = (1+|\zeta|^2)^{-2}$ is derived in Appendix~\ref{app:inf_vol_scars}.

Let us now employ the representation in Eq.~\eqref{eq:Proj_scar} to
obtain that in the large $L$ limit
\begin{equation}
\bra{\zeta}\mathcal{O}'\Pi_W\mathcal{O}\ket{\zeta} \simeq \bra{\zeta}\mathcal{O}'\ket{\zeta} \bra{\zeta}\mathcal{O}\ket{\zeta};
\label{eq:tobeproved:EM}
\end{equation}
then, substituting $\mathcal O'$ with $\mathcal O^\dagger(x,t)$, it is easy to obtain Eq.~\eqref{eq:conn_diff_proj}. To show that, we evaluate $\bra{\zeta}\mathcal{O}'\Pi_W\mathcal{O}\ket{\zeta}$ in the large $L$ limit:
\begin{equation}\label{eq:OPO_integral:EM}
\bra{\zeta}\mathcal{O}'\Pi_W\mathcal{O}\ket{\zeta} \simeq L \int_{\mathbb{C}} \frac{d^2\zeta' }{\pi} F\l|\zeta'|^2\r \bra{\zeta}\mathcal{O}'\ket{\zeta'}\bra{\zeta'}\mathcal{O}\ket{\zeta}.
\end{equation}
Recalling that $\ket{\zeta}$ is a product state and $\mathcal O$ a local operator (say, with finite support), whenever $\zeta $ and $\zeta'$ are similar but different, in the $L\to \infty$ limit 
$ \bra{\zeta'} \mathcal O \ket{\zeta} $ and $ \braket{\zeta' | \zeta}
$ scale exponentially to zero in exactly the same way, say $e^{- L/ \xi}$ with the same $\xi$. 
Therefore, the saddle point analysis in Eq.~\eqref{eq:OPO_integral:EM} is equivalent to that of 
\begin{equation}\label{eq:normaliz_coher:EM}
\bra{\zeta}\Pi_W\ket{\zeta} = L \int_{\mathbb{C}} \frac{d^2\zeta' }{\pi} F\l|\zeta'|^2\r |\braket{\zeta'|\zeta}|^2,
\end{equation}
which is more easily discussed. 
We thus expand the overlap between two close coherent states $\ket{\zeta},\ket{\zeta'}$ as
$ |\braket{\zeta'|\zeta}|^2 = \exp\l -LF_0(|\zeta|^2)|\zeta-\zeta'|^2+\dots\r$,
so that, since the integral localizes at $\zeta' \simeq \zeta$,
\begin{equation}
 \bra{\zeta} \Pi_W \ket{\zeta} \simeq \frac{F(|\zeta|^2)}{F_0 (|\zeta|^2)}
\end{equation}
that ultimately tells us that $F(|\zeta|^2) = F_0(|\zeta|^2)$.
According to what we wrote, in the large $L$ limit and for $\zeta \simeq \zeta'$, the following holds 
\begin{equation}
 \bra{\zeta} \mathcal O' \ket{\zeta'}
 \bra{\zeta'} \mathcal O \ket{\zeta} =
 \bra{\zeta} \mathcal O' \ket{\zeta}
 \bra{\zeta} \mathcal O \ket{\zeta}
 e^{- L F_0(|\zeta|^2) |\zeta - \zeta'|^2 + \ldots}
\end{equation}
and after substitution and integration in Eq.~\eqref{eq:OPO_integral:EM} we obtain Eq.~\eqref{eq:tobeproved:EM}.
With that expression, it is then possible to obtain Eq.~\eqref{eq:conn_diff_proj}.

\section{An ETH ansatz for some off-diagonal matrix elements}\label{sec:ETH}

\begin{figure}[t]
 \includegraphics[width=\columnwidth]{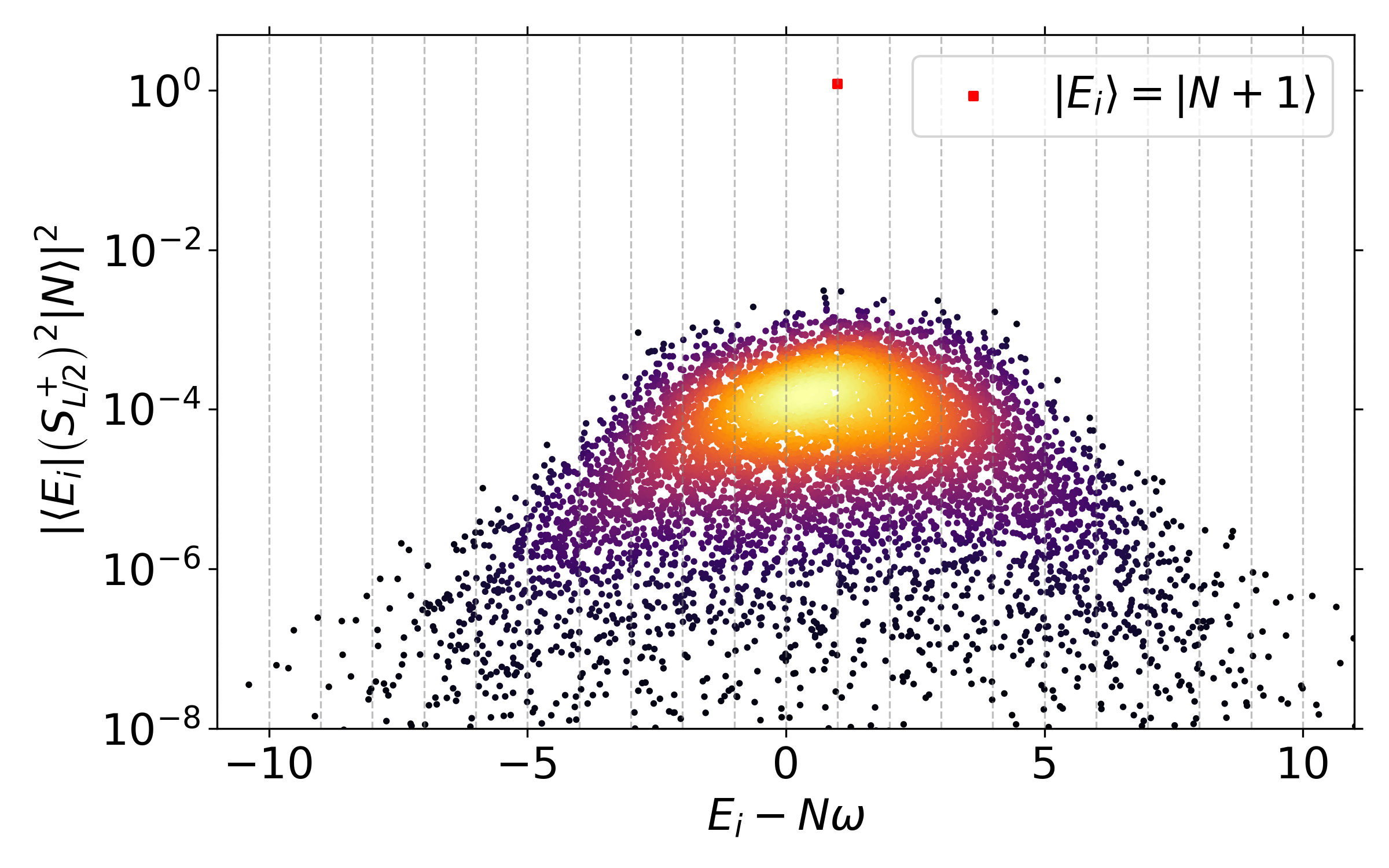}
 \caption{Off-diagonal matrix element $|\langle E_i | (S^{+}_j)^2 | N \rangle|^2$, with $j = L/2$, as a function of the energy difference $E_i - N \omega$. We choose $N=L/2$, so that the scars $\ket{N},\ket{N+1}$ have magnetization $0,2$ respectively. The red square point represents $|\langle N+1 | (S^{+}_{L/2})^2 | N \rangle|^2$, which is far larger compared to the other exponentially small overlaps. The points are calculated via exact diagonalization for $L=10$.
 The color scheme illustrates the density of points.}
 \label{fig:offdiagonal}
\end{figure}

A numerical inspection based on the exact diagonalization of a chain of length $L=10$ and reported in Fig.~\ref{fig:offdiagonal} shows that $|\bra{N+1} (S_j^+)^2 \ket{N}|^2$ is anomalously large compared to the typical value of the other ones $|\bra{E_i} (S_j^+)^2 \ket{N}|^2$.
At this point, it is tempting to speculate about the structure of the off-diagonal matrix elements and to propose a formula that extends the usual ETH ansatz, which is expected to describe $\bra {E_i} \mathcal O \ket {E_j}$ only for eigenvectors $\ket{E_i}$, $\ket{E_j}$ that do not belong to $W$.

%%%%FIG
%%%%%

We assume that every eigenstate $\ket{E_i} \notin W$ behaves in the microcanonical shell as a random vector (a similar hypothesis has been proposed in Ref.~\cite{lcm-20}). In particular, for two different eigenstates $\ket{E_i},\ket{E_j}$ in the middle of the energy spectrum that are not scarred, the corresponding off-diagonal matrix elements of local observables are expected to be exponentially small in the system size~\cite{D_Alessio_2016} as $|\bra{E_i}\mathcal{O}\ket{E_j}|^2 \sim e^{-S}$, with $S$ the microcanonical entropy extensive in $L$; conversely, their diagonal matrix elements are finite, say $\bra{E_i}\mathcal{O}\ket{E_i}\sim O(1)$ in the thermodynamic limit, whose values coincide with the associated microcanonical expectation values. This already marks an important difference with the scarred eigenstates, whose diagonal and off-diagonal matrix elements are finite in the thermodynamic limit (as shown in Appendix~\ref{app:inf_vol_scars}): specifically, $\bra{N}\mathcal{O}\ket{N+n} \sim O(1)$ in the limit $L\rightarrow \infty$ with $N/L$ and $n$ kept fixed. Thus, the only relevant case left to discuss regards the off-diagonal matrix elements between scarred and non-scarred eigenstates.

We make the general hypothesis that, for two local operators $\mathcal{O}$ and $\mathcal{O}'$: 
\begin{equation}\label{eq:Off_diag_ETH}
\begin{split}
\overline{\bra{N}\mathcal{O}'\ket{E_i} \bra{E_i}\mathcal{O}\ket{N+n}} \simeq \hspace{3.2cm} \\
\simeq e^{-S(E_i)}\times \frac{1}{2\pi} \times g^{(n)}_{\mathcal{O}',\mathcal{O}}\l \frac{N}{L}, E_i-N\omega\r,
\end{split}
\end{equation}
where $S(E_i)$ is the entropy of the associated microcanonical shell, while the bar denotes the average over the shell. We point out that $\sum_j S^z_j$, besides the energy, is an additional conserved charge: as a consequence, the ansatz \eqref{eq:Off_diag_ETH} can only be consistent if the microcanonical shell is chosen by fixing a small energy window and the magnetization sector.
In Eq.~\eqref{eq:Off_diag_ETH}, $N/L$ and $n$ are kept fixed in the infinite-size limit, and the function $g^{(n)}_{\mathcal{O}',\mathcal{O}}$ depends only on $N/L$ and on the energy difference between $\ket{E_i}$ and $\ket{N}$. 

We discuss in the following the relation between the autocorrelator of the coherent states and the matrix elements in the infinite-volume limit. We first expand it in the energy eigenbasis as in Eq.~\eqref{Eq:zeta:N:autocorr}. Then, we use $\braket{N|\zeta}\propto \zeta^N$ and that $\{|\bra \zeta N \rangle|^2\}_N$ converges to a Gaussian distribution of width $\sim \sqrt{N}$: as a consequence, the sum in \eqref{Eq:zeta:N:autocorr} is dominated by similar values $N',N$. After performing the change of variable $N' = N+n$ and employing the ansatz \eqref{eq:Off_diag_ETH}, we finally obtain
\begin{equation}
\bra{\zeta}\mathcal{O}'(t)\mathcal{O}\ket{\zeta}_c
 \simeq  \int \frac{d\omega'}{2\pi} \sum_n g^{(n)}_{\mathcal{O}',\mathcal{O}}\l \frac{N}{L}, \omega'\r \l\frac{\zeta}{|\zeta|}\r^ne^{-i\omega' t},\label{eq:autocorr_coher}
\end{equation}
where $N/L = \omega^{-1} \bra{\zeta} H \ket{\zeta}/L$ identifies the QMBS $\ket{N}$ that has the same average energy density as the coherent state $\ket{\zeta}$. 
Appendix~\ref{app:inf_vol_scars} presents a derivation of this in a more mathematical way based on saddle-point analyses.

To conclude, we specialize the prediction~\eqref{eq:autocorr_coher} to the operators $\mathcal{O} = (S^{+}_j)^2$ and $\mathcal{O}' = (S^{-}_j)^2$ for the spin-1 model in Eq.~\eqref{eq:ham}. Since $\mathcal{O}'(t) \, \mathcal Q_W \, \mathcal{O}$ commutes with the total magnetization and different scars have different magnetization, Eq.~\eqref{eq:Off_diag_ETH} can only be non-vanishing if $n=0$. 
Hence, one ends up expressing the autocorrelator of the coherent state as the Fourier transform of the function $g^{(0)}(\omega')$ through Eq.~\eqref{eq:autocorr_coher}; this implies that the origin of algebraically decaying oscillations of Eq.~\eqref{eq:autocorr_osc} is traced back to a singular behavior of $g^{(0)}(\omega')$ for $\omega' \simeq \omega$, and it is not related to the overlap between distinct scars. Similar singularities in frequency space have often been encountered in the context of transport \cite{km-63,Capizzi-24}: the novelty is that the same phenomenology is observed here, even if $\sum_j(-1)^j(S^{+}_j)^2$ is only conserved in the subspace $W$, whose dimension is exponentially small compared to the total Hilbert space.
Our numerics, limited to the exact diagonalization at a finite size $L=10$, does not allow us to collect sufficient statistics to highlight this feature.

\section{Conclusions and Outlook}\label{sec:conclusion}

We have shown that transport phenomena can emerge in systems with QMBSs due to the projected conserved charges of the scarred subspace. 
The numerical results that we presented are compatible with a superdiffusion physics with dynamical critical exponent $z=3/2$; this doe not exclude that at longer time the system crosses over towards a diffusive behaviour.
We proved that the numerically observed spreading of the autocorrelator for certain coherent states of QMBSs is mediated by the generic eigenspectrum neighboring the scars; counterintuitively, the scars do not contribute to it in the thermodynamic limit. 

Our work highlights the existence of a peculiar interplay between QMBSs and the neighboring energy spectrum: whereas in Ref.~\cite{gotta_2023} this was stated for states with momenta close to $0$ or $\pi$, here the results suggest that this should be true in the entire Brillouin zone. Highlighting this effect in other and more explicit ways, for instance, using energy filters~\cite{morettini2024energyfilter}, is an exciting outlook. 

Additionally, it would be interesting to investigate analytically the universality class of transport in similar models: in particular, in the absence of a robust theoretical explanation, we will refrain from confidently claiming superdiffusion in the long-time limit, especially due to the limitation of numerical methods in achieving reliably long times. In this regard, studies of different models with a different scar structure, such as those reported in Refs.~\cite{ktk-23,ktkk-24}, might be helpful. 
The discussion of paradigmatic kinetically-constrained models~\cite{zf-21,zbf-21} might also contribute exciting results.
It is also important to stress that a previous work has found a superdiffusive behaviour in the autocorrelation function of the infinite-temperature state of a kinetically-constrained model featuring QMBSs~\cite{ldsp-23}. However, it is not obvious that the superdiffusion that we discuss in this article has the same origin, as here it is only observed if the initial state belongs to the scarred subspace, and disappears when considering the infinite-temperature state.

Concluding, although we focused primarily on autocorrelators, which are two-point correlation functions, multipoint correlation functions of QMBSs can also be considered~\cite{Dong-22, Liang-24}.
Since they show similar patterns in space-time as the autocorrelator of the main text, it is intriguing to investigate them in connection with the general free-probability theory approach to ETH~\cite{foini_2019,foini2019eigenstate, pfk-22, fritzsch2024microcanonical, bouverotdupuis2024random, pappalardi2024eigenstate}. 

\acknowledgments

We thank M.~C.~Ba\~{n}uls, L.~Gotta, A.~Le Rose, Z.~Papic, S.~Pappalardi, M.~Serbyn and F.~M.~Surace for enlightening discussion.
LM acknowledges discussion with L. Gotta and
S. Moudgalya for a previous related work. 
LC and MF acknowledge support from Starting grant 805252 LoCoMacro. 
This work has benefited from a State grant
as part of France 2030 (QuanTEdu-France), bearing the
reference ANR-22-CMAS-0001 (GM), and is part of HQI
(www.hqi.fr) initiative, supported by France 2030 under the French National Research Agency award number
ANR-22-PNCQ-0002 (LM). This work is supported
by the ANR project LOQUST ANR-23-CE47-0006-02 (LM and LC).

\begin{appendix}

\section{QMBSs in the infinite-volume limit}\label{app:inf_vol_scars}

This Appendix presents a series of results that are relevant in order to discuss QMBSs and their properties in the infinite-volume limit.

\subsection{On the infinite-volume limit of the QMBS}

In this section, we explain how the scars in the middle of the tower induce a well-defined state for the local observables in the infinite volume limit that we characterize explicitly. We first express the scar $\ket{N}$, up to a proportionality constant, as
\be
\ket{N} \propto \frac{(J^\dagger)^N}{N!}\ket{0} = \oint \frac{d\zeta}{2\pi i \zeta} \zeta^{-N}\exp\l \zeta J^\dagger \r \ket{0},
\ee
where the integral is performed along a closed path encircling the origin $\zeta=0$ counterclockwise. This allows us to compute the expectation value of any local observable $\mathcal{O}$ via
\be\label{eq:exp_O_integral}
\begin{split}
\bra{N}\mathcal{O}\ket{N} \propto &\oint \frac{d\zeta}{2\pi i \zeta} \oint \frac{d\bar{\zeta}'}{2\pi i \bar{\zeta'}} {\bar{\zeta'}}^{-N}\zeta^{-N}\\
&\bra{0}\exp\l \bar{\zeta'}J\r\mathcal{O}\exp\l \zeta J^\dagger \r \ket{0},
\end{split}
\ee
an expression that is exact for any finite size. At this point, we focus on the limit $L\rightarrow \infty$, where $\rho\equiv N/L$ is kept fixed. To address this regime, we employ a saddle point analysis, expressing 
\be
\begin{split}
&{\bar{\zeta'}}^{-N}\zeta^{-N}\bra{0}\exp\l \bar{\zeta'}J\r\mathcal{O}\exp\l \zeta J^\dagger \r \ket{0} \asymp \\
&{\bar{\zeta'}}^{-N}\zeta^{-N}\bra{0}\exp\l \bar{\zeta'}J\r\exp\l \zeta J^\dagger \r \ket{0} \asymp \\
&\exp\left[ -L(\rho \log(\zeta \bar{\zeta'}) + f(\bar{\zeta'},\zeta))\right],
\end{split}
\ee
where we used the insertion of a local operator $\mathcal{O}$ does not change the leading exponential behavior of the overlap between two distinct (not normalized) coherent states; the function $f$ is
\be\label{eq:f_function}
f(\bar{\zeta'},\zeta)\equiv \underset{L\rightarrow \infty}{\lim} -\frac{1}{L}\log \bra{0}\exp\l \bar{\zeta'}J\r\exp\l \zeta J^\dagger \r \ket{0},
\ee
and it is well-defined since the coherent states are product states (albeit a similar statement holds for MPS). We perform a change of variable $\bar{\zeta'} = e^{i\eta}\bar{\zeta}$, we deform the integration path, and we perform the saddle point analysis of the integral
\be
\oint \frac{d\zeta}{2\pi i \zeta}\int^{\infty}_{-\infty}\frac{d\eta}{2\pi} \exp[-L(\rho\log|\zeta|^2 +i\rho \eta +f(\bar{\zeta}e^{i\eta},\zeta))].
\ee
To proceed, we choose a value of $|\zeta|$ so that the saddle for $\eta$ is at $\eta=0$, requiring
\be\label{eq:saddle_eq}
\rho + \bar{\zeta}\partial_{\bar{\zeta}}f(\bar{\zeta},\zeta) = 0
\ee
and denoting the solution by $|\zeta_\rho|$. At this point, we first integrate over $\eta$ and then we perform the contour integral over $\zeta$ at $|\zeta| = |\zeta_\rho|$. From this analysis, we learn that the integral in Eq. \eqref{eq:exp_O_integral} localizes and $\bar{\zeta'} = \zeta$ for a value of $|\zeta|=|\zeta_\rho|$ that is fixed by Eq. \eqref{eq:saddle_eq}. In conclusion, we obtain
\be\label{eq:incoher_superpos}
\bra{N}\mathcal{O}\ket{N} \underset{L\rightarrow \infty}{=} \oint_{|\zeta| = |\zeta_\rho|}\frac{d\zeta}{2\pi i \zeta} \bra{\zeta}\mathcal{O}\ket{\zeta}.
\ee
Such a general result, valid for any local observable $\mathcal{O}$, expresses an important property of the infinite volume limit of the scars; for instance, while at any finite size the states $\ket{N}$ are pure, they become locally indistinguishable from incoherent superpositions of distinct (except for the trivial cases $\zeta=0,\infty$ corresponding to $N/L =0,1$ respectively) coherent states when the infinite volume limit is considered: specifically, recalling that the phase $\zeta$ evolves periodically in time, we discover that is not possible distinguish locally the time averages over coherent states from the expectation values over the scar, and we equivalently express Eq. \eqref{eq:incoher_superpos} as
\be
\bra{N}\mathcal{O}\ket{N} \underset{L\rightarrow \infty}{=} \omega\int^{\frac{2\pi}{\omega}}_0 \frac{dt}{2\pi}\bra{\zeta}\mathcal{O}(t)\ket{\zeta}, \quad \zeta = |\zeta_\rho|.
\ee
Further, we observe that an equivalent way to compute explicitly $|\zeta_\rho|$, beside Eq. \eqref{eq:saddle_eq}, amounts to compare the value of magnetization of $\ket{\zeta}$ and $\ket{N}$: in particular, we choose $\mathcal{O} = S^z_j$ and we solve Eq. \eqref{eq:incoher_superpos}, where the integral over $\zeta$ is constant, to obtain the value of $|\zeta_\rho|$.

A striking consequence of Eq. \eqref{eq:incoher_superpos} is the presence of long-range correlations, both in space and time, for the scars $\ket{N}$, a property that has been recognized in Ref. \cite{isx-19} for the local operator $\mathcal{O}=(S^+_j)^2$. Specifically, for any pair of local operators $\mathcal{O},\mathcal{O}'$ we compute the connected correlator in the limit of large $x$ as
\be\label{eq:LRO_space}
\begin{split}
&\bra{N}\mathcal{O}(x)\mathcal{O}'\ket{N} -\bra{N}\mathcal{O}(x)\ket{N}\bra{N}\mathcal{O}'\ket{N} =\\
&\overline{\bra{\zeta}\mathcal{O}(x)\ket{\zeta}\bra{\zeta}\mathcal{O}'\ket{\zeta}} - \overline{\bra{\zeta}\mathcal{O}(x)\ket{\zeta}}\times \overline{\bra{\zeta}\mathcal{O}'\ket{\zeta}},
\end{split}
\ee
where $\bra{\zeta}\mathcal{O}(x)\mathcal{O}'\ket{\zeta} \simeq \bra{\zeta}\mathcal{O}(x)\ket{\zeta}\bra{\zeta}\mathcal{O}'\ket{\zeta}$ has been employed. Here $\overline{(\dots)}$ denotes the integral over the coherent states at given $|\zeta|=|\zeta_\rho|$ expressed by the r.h.s. of Eq. \eqref{eq:incoher_superpos}. For a generic choice of the operators, as for $\mathcal{O}'=(S^{+}_j)^2$ and $\mathcal{O}=(S^{-}_j)^2$, Eq. \eqref{eq:LRO_space} is non-vanishing and therefore long-range correlations are present in space. A similar analysis can be performed for the temporal correlation, replacing $\mathcal{O}(x)\rightarrow \mathcal{O}(t)$ in Eq. \eqref{eq:LRO_space}: the only technical issue is that we are not able to show rigorously that $\underset{t\rightarrow \infty}{\lim}\bra{\zeta}\mathcal{O}(t)\mathcal{O}'\ket{\zeta}_c =0$, albeit our numerics is consistent with this property.

As a final technical point, we report the explicit expression of the function $f(\bar{\zeta'},\zeta)$ appearing in Eq. \eqref{eq:f_function}. Starting from the definition of $J^\dagger$ in Eq. \eqref{eq:j_dag}, we express
\be
\exp(\zeta J^\dagger)\ket{0} = \prod_j\l (-1)^j \zeta \ket{\uparrow}_j +\ket{\downarrow}_j\r,
\ee
which gives directly $\bra{0}\exp\l \bar{\zeta'}J\r\exp\l \zeta J^\dagger \r \ket{0} = (1+\bar{\zeta'}\zeta)^L$, and, from Eq. \eqref{eq:f_function}, we identify
\be\label{eq:f_function_1}
f(\bar{\zeta'},\zeta) = -\log(1+\bar{\zeta'}\zeta).
\ee
Therefore, the saddle-point relation in Eq. \eqref{eq:saddle_eq} is
\be
\rho = \frac{|\zeta_\rho|^2}{1+|\zeta_\rho|^2},
\ee
and it has a unique solution for any $\rho \in [0,1]$; this is consistent with the range of values $N/L \in [0,1]$ in Eq. \eqref{eq:j_dag}.

\subsection{Off-diagonal matrix elements between scars}

In this section, we show that the off-diagonal matrix element of a local observable $\mathcal{O}$ between distinct scars $\ket{N},\ket{N+n}$, that we denote by $\bra{N}\mathcal{O}\ket{N+n}$ has a finite value in the limit of $L\rightarrow \infty$ with $N/L$ and $n$; its specific value, which can be zero in principle, depends clearly on $\mathcal{O}$. We first express, up to a normalization constant
\be
\begin{split}
\bra{N}\mathcal{O}\ket{N+n} \propto &\oint \frac{d\zeta}{2\pi i \zeta} \oint \frac{d\bar{\zeta}'}{2\pi i \bar{\zeta'}} {\bar{\zeta'}}^{-N}\zeta^{-(N+n)}\\
&\bra{0}\exp\l \bar{\zeta'}J\r\mathcal{O}\exp\l \zeta J^\dagger \r \ket{0},
\end{split}
\ee
which amounts to replace $ \zeta^{-N}\rightarrow \zeta^{-(N+n)}$ in Eq. \eqref{eq:exp_O_integral}. The calculation is analogous to that in Eq. \eqref{eq:exp_O_integral}, since the saddle point analysis is not modified by the presence of $n$ (that is kept fixed). We perform similar steps, we identify the saddle from Eq. \eqref{eq:saddle_eq}, and eventually, taking into account the correct normalization of the state, we obtain
\be\label{eq:On}
\bra{N}\mathcal{O}\ket{N+n} \underset{L\rightarrow \infty}{=} \oint_{|\zeta| = |\zeta_\rho|}\frac{d\zeta}{2\pi i \zeta} \left(\frac{\zeta}{|\zeta|}\right)^{-n}\bra{\zeta}\mathcal{O}\ket{\zeta}
\ee
that generalizes Eq. \eqref{eq:incoher_superpos} for $n \neq 0$. Similarly, applying the change of variables $\zeta = e^{-i\omega t}|\zeta_\rho|$, we express equivalently
\be\label{eq:off_diag_integ_t}
\bra{N}\mathcal{O}\ket{N+n} \underset{L\rightarrow \infty}{=} \omega\int^{\frac{2\pi}{\omega}}_0 \frac{dt}{2\pi}e^{in\omega t}\bra{\zeta}\mathcal{O}(t)\ket{\zeta},
\ee
with $\quad \zeta = |\zeta_\rho|$, that is the main result of this section. Finally, we invert the relation above and we find
\be
\bra{\zeta}\mathcal{O}(t)\ket{\zeta} = \sum_n \bra{N}\mathcal{O}\ket{N+n}e^{-in\omega t}, \quad \zeta = |\zeta_\rho|.
\ee

As a simple example, we observe that for $\mathcal{O}=(S^-_j)^2$ it holds $\bra{\zeta}\mathcal{O}(t)\ket{\zeta} = \bra{\zeta}\mathcal{O}\ket{\zeta}e^{-i\omega t}$, and therefore Eq. \eqref{eq:off_diag_integ_t} is non-vanishing when $n=1$ only; also,  for $\mathcal{O}=(S^z_j)$, $\bra{\zeta}\mathcal{O}(t)\ket{\zeta} = \bra{\zeta}\mathcal{O}\ket{\zeta}$ and Eq. \eqref{eq:off_diag_integ_t} vanishes for any $n$ but $n=0$. In general, one expects that the matrix element in Eq. \eqref{eq:off_diag_integ_t} vanishes, or decays fast, for large $|n|$: this corresponds to the impossibility of connecting states with a large difference in magnetization (or energy) via a local operator. Such a property can be proven rigorously for local observables starting from Eq. \eqref{eq:off_diag_integ_t}. In particular, one notices that $\bra{\zeta}\mathcal{O}(t)\ket{\zeta}$ is a smooth function of $t$, as a consequence of locality (we refer the reader to Refs. \cite{br-v1,br-v2} for details), and it is periodic with period $2\pi/\omega$; therefore, from standard mathematical results, its discrete Fourier series goes to zero faster than any power-law as $|n|\rightarrow \infty$.

\subsection{Irrelevance of the scar projector inside connected correlator}

In this section, we will prove the relation \eqref{eq:conn_diff_proj}, that expresses the irrelevance of the scar inside the connected correlators of coherent states. We will first give an explicit formula of the projector $\Pi_W$ in the limit of large $L$, reported in Eq. \eqref{eq:Proj_scar} of the main text. This is standard textbook material \cite{Radcliffe-71}, and it amounts to express the identity operator in terms of coherent states in a spin-$S$ representation of $SU(2)$ for large $S$ (here, $2S=L$). For completeness, and for later purposes, we provide here a concise derivation using saddle-point techniques.

We start from the ansatz \eqref{eq:Proj_scar} for some function $F$ which depends on $\zeta'$ through its modulus. This ansatz is motivated, since, after expressing $\ket{\zeta}$ in terms of the scars $\ket{N}$ and performing the integral over the complex plane, a sum of terms proportional to $\ketbra{N}{N}$ will be generated, that is, off-diagonal matrix elements are never generated: the functional form of $F$ can be fixed by the requirement $\bra{\zeta}\Pi_W\ket{\zeta}=1$. In particular, we first expand the overlap between two close coherent states $\ket{\zeta},\ket{\zeta'}$ as
\be
|\braket{\zeta'|\zeta}|^2 = \exp\l -LF_0(|\zeta|^2)|\zeta-\zeta'|^2+\dots\r.
\ee
Then, we compute
\be\label{eq:normaliz_coher}
\bra{\zeta}\Pi_W\ket{\zeta} = L \int_{\mathbb{C}} \frac{d\zeta' d\bar{\zeta'}}{\pi} F\l|\zeta'|^2\r |\braket{\zeta'|\zeta}|^2,
\ee
using the saddle-point approximation around $\zeta'\simeq \zeta$ for large $L$, and finally we obtain
\be
\bra{\zeta}\Pi_W\ket{\zeta} = \frac{F\l|\zeta|^2\r}{F_0\l|\zeta|^2\r},
\ee
meaning that $F=F_0$. In our case, the function $F_0$ can be determined from Eq. \eqref{eq:f_function_1}, since
\be\begin{split}
&|\braket{\zeta'|\zeta}|^2 = \\
&\frac{|\bra{0}\exp(\bar{\zeta'}J)\exp(\zeta J^\dagger)\ket{0}|^2}{\bra{0}\exp(\bar{\zeta'}J)\exp(\zeta'J^\dagger)\ket{0}\bra{0}\exp(\bar{\zeta}J)\exp(\zeta J^\dagger)\ket{0}} \asymp \\
&\exp\l -L(f(\bar{\zeta'},\zeta)+f(\bar{\zeta},\zeta')-f(\bar{\zeta'},\zeta')-f(\bar{\zeta},\zeta))\r.
\end{split}
\ee
Thus, we expand $\zeta' = \zeta e^{z}$ for small $z$, and using the Taylor expansion
\be\begin{split}
\log(1+|\zeta|^2 e^{z}) = &\log(1+|\zeta|^2 ) +\\ &\frac{|\zeta|^2}{(1+|\zeta|^2)}z+ \frac{|\zeta|^2}{2(1+|\zeta|^2)^2}z^2+\dots
\end{split}
\ee
we compute
\be
\begin{split}
|\braket{\zeta'|\zeta}|^2 \simeq &\exp\l -L\frac{|\zeta|^2}{(1+|\zeta|^2)^2}z\bar{z}\r \simeq \\
&\exp\l -L\frac{1}{(1+|\zeta|^2)^2}|\zeta-\zeta'|^2\r,
\end{split}
\ee
and we identify $F(|\zeta|^2) = \frac{1}{(1+|\zeta|^2)^2}$.

The same technique can be employed to evaluate $\bra{\zeta}\mathcal{O}\Pi_W\mathcal{O}'\ket{\zeta}$ in the large $L$ limit. We first use the representation \eqref{eq:Proj_scar} of $\Pi_W$ and we compute
\be\label{eq:OPO_integral}
\bra{\zeta}\mathcal{O}'\Pi_W\mathcal{O}\ket{\zeta} \simeq L \int_{\mathbb{C}} \frac{d\zeta' d\bar{\zeta'}}{\pi} F\l|\zeta'|^2\r \bra{\zeta}\mathcal{O}'\ket{\zeta'}\bra{\zeta'}\mathcal{O}\ket{\zeta}.
\ee
Then, we observe that, for large $L$, $\bra{\zeta'}\mathcal{O}\ket{\zeta} \asymp \braket{\zeta'|\zeta}$, that is the insertion of a local operator $\mathcal{O}$ does not modify the exponential behavior in $L$ of the overlap. Therefore, the saddle point analysis in Eq. \eqref{eq:OPO_integral} is equivalent to that of Eq. \eqref{eq:normaliz_coher}, the integral localizes at $\zeta' \simeq \zeta$, and the leading term of Eq. \eqref{eq:OPO_integral} is
\be
\bra{\zeta}\mathcal{O}'\Pi_W\mathcal{O}\ket{\zeta} \simeq \bra{\zeta}\mathcal{O}'\ket{\zeta} \bra{\zeta}\mathcal{O}\ket{\zeta},
\ee
as anticipated in Eq. \eqref{eq:conn_diff_proj}. The previous expression becomes exact in the limit $L\rightarrow \infty$: clearly, at a finite size, corrections are present and they mostly come from a proper evaluation of the integral besides the saddle-point approximation. However, these subleading effects are beyond our purpose.

\subsection{Relation between autocorrelators of scars and coherent states}

In this section, we relate the connected correlations of the coherent state to that of the scars, which is useful to get Eq. \eqref{eq:autocorr_coher}. Informally, that can be obtained from Eq. \eqref{eq:On} regarding $\Pi_W$ as a local operator and replacing $\Pi_W \rightarrow \ketbra{\zeta}{\zeta}$ inside the correlators of the coherent state $\ket{\zeta}$: while the final result is correct, it is not justified, since, strictly speaking, $\Pi_W$ is not local. To overcome this technical issue, we give below a rigorous proof which make use of the results of the previous appendices. 

We consider two generic local observables $\mathcal{O},\mathcal{O}'$, and we compute (for $N/L,n$ fixed)
\be\label{eq:OPO_n}
\begin{split}
&\bra{N}\mathcal{O}'\Pi_W\mathcal{O}\ket{N+n} = \\
&\sum_{n' \in \mathbb{Z}}\bra{N}\mathcal{O}'\ket{N+n'}\bra{N+n'}\mathcal{O}\ket{N+n} \underset{L\rightarrow \infty}{=} \\
&\oint_{|\zeta| = |\zeta_\rho|}\frac{d\zeta}{2\pi i \zeta} 
\oint_{|\zeta'| = |\zeta_\rho|}\frac{d\zeta'}{2\pi i \zeta'} \bra{\zeta}\mathcal{O}'\ket{\zeta}\bra{\zeta'}\mathcal{O}\ket{\zeta'}\\
&\l\frac{\zeta'}{|\zeta'|}\r^{-n}\sum_{n' \in \mathbb{Z}}\l\frac{\zeta'}{\zeta}\r^{-n'}.
\end{split}
\ee
We change variables $\zeta' = \zeta e^{i\eta}$,  we replace
\be
\oint\frac{d\zeta'}{2\pi i \zeta'} \rightarrow \int^{+\pi}_{-\pi} \frac{d\eta}{2\pi},
\ee
and we employ the relation 
\be
\sum_{n' \in \mathbb{Z}} e^{in'\eta} = 2\pi \delta(\eta),
\ee
valid in the distributional sense over the smooth functions of $\eta$. Thus, Eq. \eqref{eq:OPO_n} gives
\be
\begin{split}
&\bra{N}\mathcal{O}'\Pi_W\mathcal{O}\ket{N+n} \underset{L\rightarrow \infty}{=}\\
&\oint_{|\zeta| = |\zeta_\rho|}\frac{d\zeta}{2\pi i \zeta} \bra{\zeta}\mathcal{O}'\ket{\zeta}\bra{\zeta}\mathcal{O}\ket{\zeta}\l\frac{\zeta}{|\zeta|}\r^{-n},
\end{split}
\ee
and using Eq. \eqref{eq:On}, we obtain
\be
\begin{split}
&\bra{N}\mathcal{O}'(1-\Pi_W)\mathcal{O}\ket{N+n} \underset{L\rightarrow \infty}{=} \\
&\oint_{|\zeta| = |\zeta_\rho|}\frac{d\zeta}{2\pi i \zeta} \bra{\zeta}\mathcal{O}'\mathcal{O}\ket{\zeta}_c \l\frac{\zeta}{|\zeta|}\r^{-n}.
\end{split}
\ee
We invert this relation, via Fourier transform, and we finally get
\be
\bra{\zeta}\mathcal{O}'\mathcal{O}\ket{\zeta}_c \underset{L\rightarrow \infty}{=} \sum_{n\in \mathbb{Z}} \l\frac{\zeta}{|\zeta|}\r^{n} \bra{N}\mathcal{O}'(1-\Pi_W)\mathcal{O}\ket{N+n}.
\ee
This is the result in Eq. \eqref{eq:autocorr_coher} after the replacement $\mathcal{O}'\rightarrow \mathcal{O}'(t)$ and making use of the ETH ansatz \eqref{eq:Off_diag_ETH}.

\section{Additional numerical details}
\label{App:Numerics}

%%%%
\begin{figure}[ht]
 \includegraphics[scale=0.5]{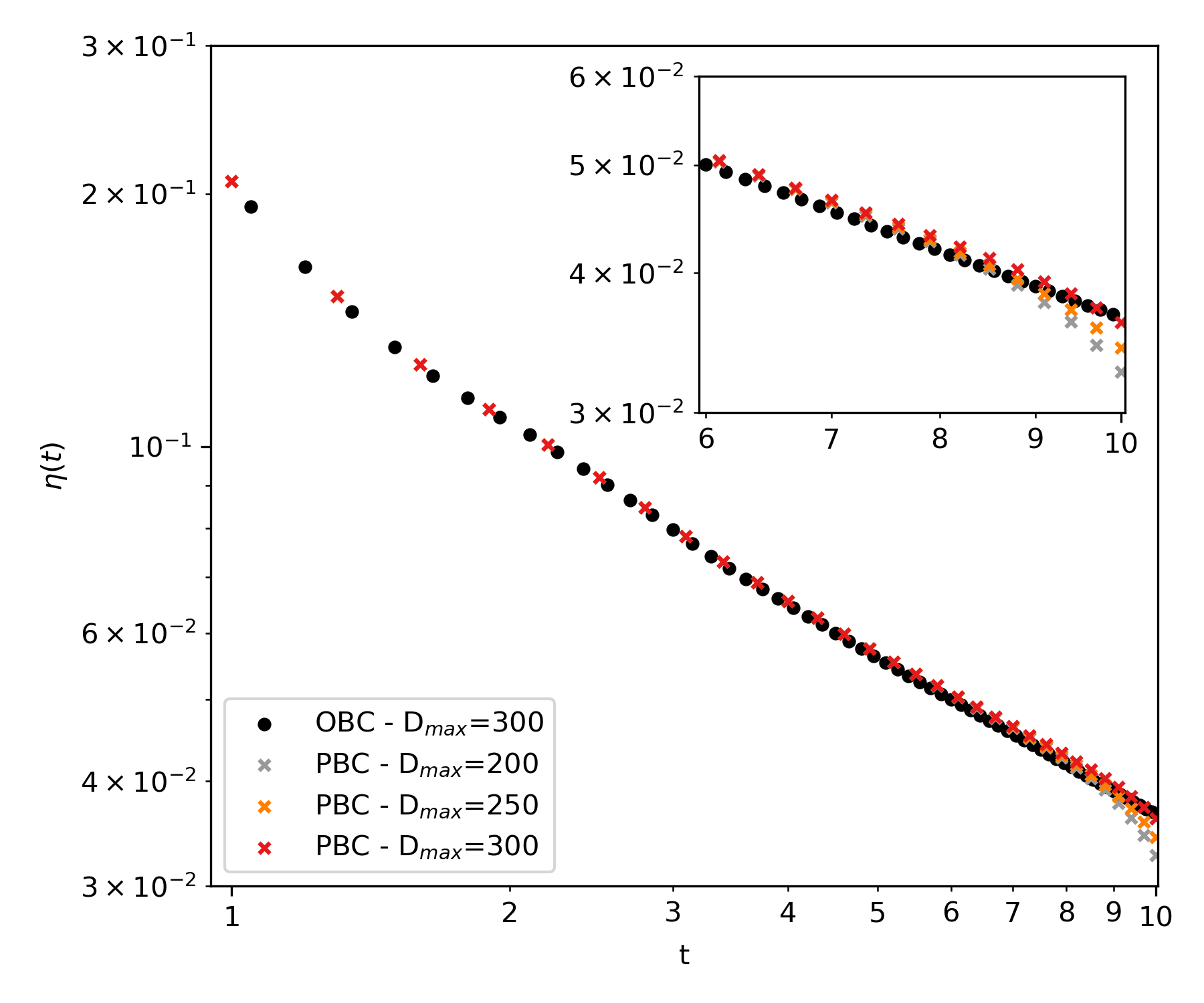}
 \caption{
 $\eta(t)$ as a function of time $t$. We compare the simulation of the main text (OBC) with those of the new algorithm (PBC). The inset highlights the discrepancies, observed for different bond dimensions, relevant at later times.
 }
 \label{fig:appendix}
\end{figure}
%%%%%

In this section, we show an additional numerical test, performed with a different algorithm, to probe the autocorrelator considered in the main text. We have exploited some symmetries of the system to reduce the numerical complexity, focusing on periodic boundary conditions (rather than open, as in the main text): this is an additional check for the irrelevance of boundary conditions on the timescales considered. We first discuss the symmetries, and then we give the details of the new algorithm, comparing the results with those of the main text.

We use the conservation of magnetization, and we split the Hamiltonian as $H = H_0+h\sum_j S^z_j$ with $\ket{\zeta}$ satisfying $e^{-iH_0t}\ket{\zeta} = \ket{\zeta}$ (additional phases, that can be absorbed in an additive constant for $H_0$, are irrelevant for our purposes). In particular, the origin of the oscillations discussed in the main test traces back to the term of the Hamiltonian proportional to the magnetization and it can be factorized out from the dynamics. Specifically, one can easily verify that
\be
(S^{-}_{j})^2(t) = e^{iHt}(S^{-}_{j})^2 e^{-iHt} = e^{-i2ht} e^{iH_0t}(S^{-}_{j})^2 e^{-iH_0t}
\ee
and, from this point onward, we will focus on $h=0$ without loss of generality. Under this assumption, we write the autocorrelator of $(S^+)^2$ as
\be\label{eq:rep_autocorr}
\bra{\zeta}(S^-_{j})^2(t)(S^+_{j'})^2(0)\ket{\zeta}_c = \bra{\zeta}(S^-_{j})^2(t/2)(S^+_{j'})^2(-t/2)\ket{\zeta}_c.
\ee
The suggestive representation is particularly useful for numerics (see also Ref. \cite{kbm-13,kk-16}): for instance, we will store the MPS approximating the ket $(S^+_{j'})^2(-t/2)\ket{\zeta}$ for a given reference position, say $j'=L/2$; then, using translational invariance and the other symmetries, we generate the bra $\bra{\zeta}(S^-_{j})^2(t/2)$ to finally reconstruct the connected correlation function at any point performing the scalar products with the ket. To do so, some minor technical caveats arise, and we explain them below. First, by performing the adjoint followed by complex conjugation (in the $z$ basis) on the aforementioned ket, we obtain
\be
[((S^+_{j'})^2(-t/2)\ket{\zeta})^{\dagger}]^{*} = \bra{\zeta^*}(S^-_{j'})^2(t/2),
\ee
where the property (time-reversal invariance) $H=H^*$ has been employed. In the following, we will consider $\zeta$ real, so that $\zeta = \zeta^*$, and we choose $\zeta=1$: while this is different with respect to the choice $\zeta=-i$ of the main text, one can easily show, using the conservation of the magnetization, that the autocorrelator \eqref{eq:rep_autocorr} does not depend on the phase of $\zeta$ explicitly. Lastly, given $T$ the translation operator of a single site, we find
\be
 \bra{\zeta (-1)^x}(S^-_{j'+x})^2(t/2) = \bra{\zeta}(S^-_{j'})^2(t/2) T^{x},
\ee
coming from the transformation $\ket{\zeta}\rightarrow \ket{-\zeta}$ under one-site translation: this causes an issue whenever the translation of an odd number of sites has to be performed. However, we can overcome the problem since $\bra{\zeta}$ can be related to $\bra{-\zeta}$ using a rotation along the $z$-axis, and, after straightforward algebra, we express
\be
\bra{\zeta }(S^-_{j'+x})^2(t/2)  = -i^L \bra{\zeta}(S^-_{j'})^2(t/2) T^{x}e^{i\pi/2 \sum_j S^z_j},
\ee
with $x$ odd. Putting these relations together, one can efficiently reconstruct \eqref{eq:rep_autocorr} calculating the aforementioned bras and kets and their respective scalar product. For the sake of completeness we remark that, so far, the only source of discrepancy with respect to the main text comes from the boundary conditions: however, as previously stressed, we expect that they do not play a significant role for the protocol under analysis, as a consequence of the locality of the interactions.

%Regarding the techniques for the unitary evolution, two different approaches have been used. In the main text, we use the Time-Dependent Variational Principle (TDVP). This method can become unreliable if the bond dimension of the MPS is too small. To mitigate this issue, we first apply the two-site TDVP with a very high precision (cutoff $\epsilon=10^{-15}$), allowing the dimension of the bond to increase dynamically and fast enough to represent the evolved state correctly. Once the maximum bond dimension ($\text{D}_{max}=300$) is reached, we switch to one-site TDVP, which is computationally less expensive. The time step and sweep number used are $\delta t= 0.05$ and $n_{sweeps}=5$, respectively.

We now describe the technique employed for the unitary evolution, that allows to represent efficiently $(S^+_{j'})^2(-t/2)\ket{\zeta}$ as an MPS. We combine Time-evolving block decimation (TEBD) and TDVP~\cite{Schollwock2019, Haegeman_2011, Haegeman2016}. We remark that TEBD is well-suited for evolving initial product states, but it struggles with periodic boundary conditions. To overcome these limitations, we use TEBD with high accuracy up to a fixed time, ensuring that the MPS reaches a sufficiently large bond dimension. At this point, we switch to two-site TDVP. Finally, once the bond dimension reaches a given threshold $\text{D}_{\text{max}}$, we perform the one-site TDVP~\cite{White2021, Goto_2019}.
The parameters used for the TEBD are $\left[ \delta t, \epsilon, t_f \right] = \left[ 0.02, 10^{-14}, 0.2 \right]$, where $t_f$ is the time after which we apply TDVP. For the TDVP we use $\left[ \delta t, \epsilon, n_{sweeps}\right]=\left[0.1, 10^{-12}, 1\right]$. We compare the algorithm described above, for periodic boundary conditions (PBC), and that of the main text with open boundary conditions (OBC); in particular, we have kept the same values for the parameters $J,h,D,J_3$ and the size is $L=60$. In Fig.~\ref{fig:appendix} we plot the quantity $\eta(t)$ for different bond dimensions: an inset highlights the discrepancies that are negligible for $t\lesssim 8$.

\end{appendix}

\twocolumngrid
\bibliography{bibliography.bib}

\end{document}